\documentclass{iau-2022}

\usepackage{amsmath}
\usepackage{graphicx}
\usepackage{multirow}

\begin{document}

\lefttitle{Kyungmin Kim}
\righttitle{Search for Microlensing Signature in GWs from Binary Black Hole Events}

\jnlPage{1}{2}
\jnlDoiYr{2022}
\doival{}

\volno{xxx}
\aopheadtitle{Proceedings IAU Symposium}
\editors{A. Mahabal,  C. Fluke \&  J. McIver, eds.}

\title{Search for Microlensing Signature in Gravitational Waves from Binary Black Hole Events}

\author{Kyungmin Kim}
\affiliation{Department of Physics, Ewha Womans University, Seoul 03760, South Korea}

\begin{abstract} 
In a recent search~\citep{Kim:2022lex}, we looked for microlensing signature in gravitational waves from spectrograms of the binary black hole events in the first and second gravitational-wave transient catalogs. For the search, we have implemented a deep learning-based method~\citep{Kim:2020xkm} and figured out that one event, GW190707\_093326, out of forty-six events, is classified into the lensed class. However, upon estimating the $p$-value of this event, we observed that the uncertainty of the $p$-value still includes the possibility of the event being unlensed. Therefore, we concluded that no significant evidence of beating patterns from the evaluated binary black hole events has found from the search. For a consequence study, we discuss the distinguishability between microlensed GWs and the signal from precessing black hole binaries.
\end{abstract}

\begin{keywords}
Gravitational waves, Gravitational microlensing
\end{keywords}

\maketitle

\section{Previous Searches}
Similar to light, gravitational waves (GWs) can be lensed when they propagate near massive astrophysical objects. To date, many searches have tried to look for the signature of the gravitational lensing of GWs (GW lensing, hereafter)~\citep[and references therein]{Kim:2022lex}.
However, there has been no widely accepted detection thus far.

Alternatively, in a recent search~\citep{Kim:2022lex}, we have revisited the forty-six BBH events analysed in \cite{Hannuksela:2019kle, LIGOScientific:2021izm} and evaluated them with the deep learning-based classification method built in an earlier work~\citep{Kim:2020xkm}: We implement one of the state-of-the-art deep learning
models, VGG-19~\citep{Simonyan14c} by setting the batch size to be 128, the maximum training epoch to be 10 with adopting the Adam optimizer \citep{2014ADAM} and the cross-entropy loss for the training. We newly introducing ensemble learning to mitigate possible biased prediction from a single learner. On top of that, we have built a new hierarchical classification scheme for the evaluation of BBH signals. One can refer \cite{Kim:2020xkm, Kim:2022lex} and the attached supplementary material for more details of utilizing the deep learning model. 

From the search, we figured out only one event, GW190707\_093326, out of forty-six events is primarily classified as lensed. To verify the result, we estimated the $p$-value for the median probability $\bar{r}=0.984^{+0.012}_{-0.342}$ and observed that the uncertainty of the estimated $p$-value, ranging $0 \lesssim p \lesssim 0.1$, is still containing $p \geq 0.05$ where the unlensed hypothesis is true. Therefore, we have concluded that the search has found no significant evidence of beating patterns from the evaluated BBH events.

\section{Can we distinguish microlensed GWs from the signal of precessing BBH mergers?}
\begin{figure}[t!]
    \centering
    \includegraphics[width=.82\linewidth]{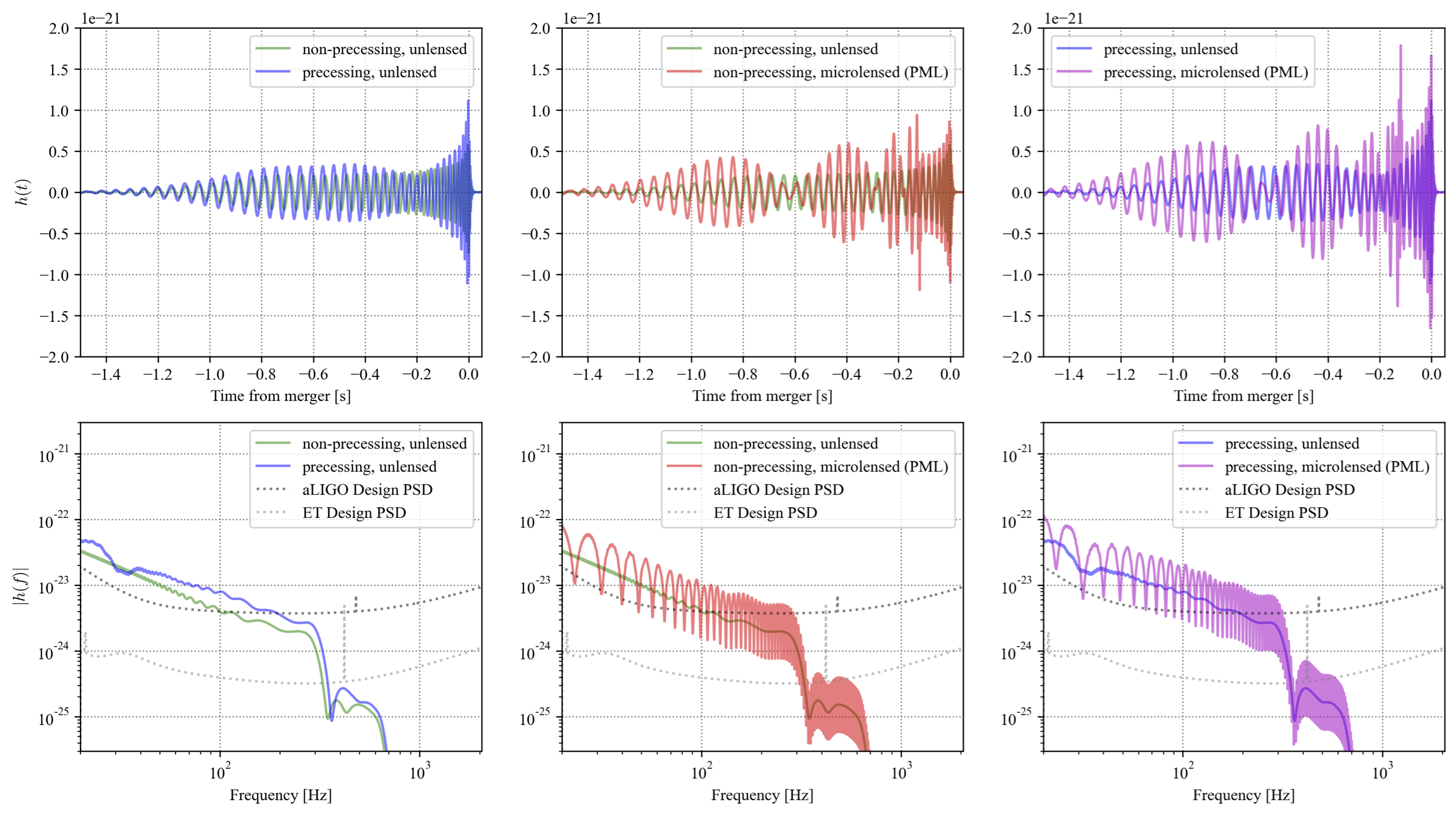}
    \caption{Example waveforms showing how different a microlensed GW signal against the signal of a precessing BBH merger. We assume the same component mass ($m_1=m_2=30M_\odot$) and the same distance to the source ($500\textrm{Mpc}$) for both cases. For the generation of the signal from precessing BBH merger, we set arbitrary spin vectors for each component black hole. For microlensed signals, we adopt the point-mass lens model in the geometrical optics limit with the source position parameter $y=0.3$ and the redshifted lens mass $M_{lz}=10^4M_\odot$. (see \cite{takahashi:2003apj} for details of lensing configuration and formulation.)}
    \label{fig}
\end{figure}

In the two previous works, we have supposed the microlensing of non-precessing GWs for convenience. However, it is known that precessing BBH systems can introduce modulation in the waveform which looks quite similar to the beating pattern shown in the waveform of microlensed GWs. In Fig.~\ref{fig}, we present example waveforms corresponding to the cases: The left panels show how the waveforms from non-precessing (green) and precessing (blue) look different in the time domain (top) and frequency domain (bottom). In the middle and right panels, we provide waveforms of microlensed signals of non-precessing and precessing BBH mergers, respectively. From the comparison between unlensed precessing waveform and lensed non-precessing waveform in the time domain (the top panels of the left and middle panels, respectively), we can observe the modulations shown in the inspiral phase are rather different. In the frequency domain, the modulation from the precession appears as a smooth sinusoidal oscillation while the beating pattern by the microlensing produces arch-shaped patterns. Moreover, in the time-domain waveform, we can see there are two peaks between [-0.2, 0] seconds for the lensed signal. However, when we compare the middle and right panels, the GW microlensing from the same lensing configuration introduces very similar modulations which hinders to identify whether it came from non-precessing or precessing binaries. 

From this preliminary qualitative study, it looks feasible to certainly distinguish the microlensing-induced beating patterns from the precession-induced modulation. However, it will be quite challenging to distinguish whether the signal is from non-pressing or precessing binaries when both cases are microlensed. We expect exploiting a machine learning method such as the distance metric learning will be a possible method for an exploration of the distinguishability even in noisy data in terms of the similarity between the target of interest.

\end{document}